\documentclass[%
 reprint,
 amsmath,amssymb,
 aps,
]{revtex4-2}

\usepackage{graphicx}
\usepackage{dcolumn}
\usepackage{bm}
\usepackage{hyperref}
\usepackage[mathlines]{lineno}
\usepackage{bm}
\usepackage{float}
\usepackage{cleveref}
\usepackage{wasysym}
\usepackage[makeroom]{cancel}
\usepackage{scalerel,amssymb}

\begin{document}
   \sloppy 
\preprint{APS/123-QED}
\allowdisplaybreaks
\title{A simple, general criterion for onset of disclination disorder on curved surfaces}

\author{Siddhansh Agarwal}
\author{Sascha Hilgenfeldt}%
 \email{sascha@illinois.edu}
\affiliation{Mechanical Science and Engineering, University of Illinois, Urbana-Champaign, Illinois 61801, USA}%

\date{\today}

\begin{abstract}
Determining the positions of lattice defects on elastic surfaces with Gaussian curvature is a non-trivial task of mechanical energy optimization, particularly for surfaces with boundaries. We introduce a simple way to predict the onset of disclination disorder from the shape of bounded surfaces. The criterion fixes the value of a weighted integral Gaussian curvature to a universal constant and proves accurate across a great variety of shapes, even when previously suggested criteria fail. It is an easy avenue to improved  understanding of the limitations to crystalline order in many materials.
\end{abstract}

\maketitle

Crystalline domain systems with intrinsic curvature that are governed by the minimization of an interaction energy are commonly found in nature. Extensive studies on viral capsids, vesicles and other curved ``soft" crystals have revealed an interplay between curvature, crystalline order, and overall structure in the system \cite{meng2014elastic,kohler2016stress,bausch2003grain,drenckhan2004demonstration,sknepnek2012buckling,yong2013elastic}. As an example, initially spherical viral capsids are known to buckle into a faceted geometry when the F{\"o}ppl-von K\'arm\'an (FvK) number exceeds a threshold value \cite{lidmar2003virus,vsiber2006buckling}. 

Unlike in flat 2D space where interacting particles can pack in triangular lattices, finite Gaussian curvature $K_G$ introduces geometric frustration. This concept is 
quantified by the Euler theorem, requiring a total topological charge on the lattice of 
$Q = \sum_{i=1}^V q_i = 6\chi$, where $q_i=6-c_i$ using $c_i$ for the coordination number of the $i$th vertex, and the Euler characteristic $\chi$.
These disclinations contribute inevitably to elastic energy of the shell \cite{nelson2002defects}. Finding the positioning of defects that minimizes mechanical energy in more general scenarios is a formidable task of considerable recent interest \cite{vitelli2004defect,bowick2002crystalline,giomi2007crystalline,perez1997influence,mughal2014packing,burke2015role}.
Even on developable ($K_G=0$) surfaces, defects can be intricately connected with the crystalline structure if polydispersity is present, as shown in recent studies of the onset of packing defects in the compound eye of \textit{Drosophila} fly pupae \cite{kim2016hexagonal}. However, in later developmental stages the retinal tissue develops Gaussian curvature, and the question of energetically optimal defect positioning must be asked anew.
\begin{figure}
	\centering
		\includegraphics[width=0.45\textwidth]{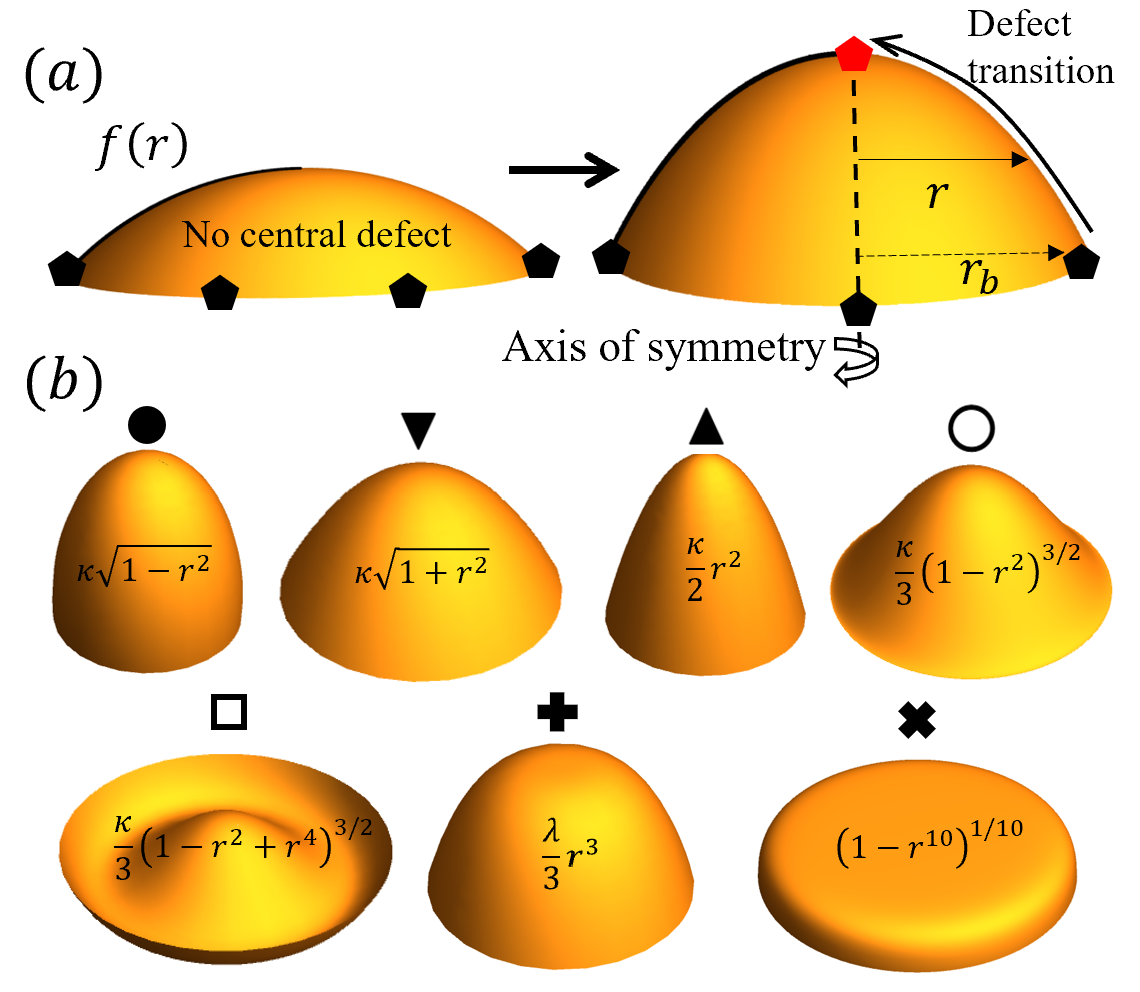}
	\caption{(a) A single disclination defect at the apex becomes energetically favourable for large enough central curvature $\kappa$ and/or cap extent $r_b$;
	(b) Sample families of rotationally symmetric cap surfaces with their respective parametrizations $f(r)$; the first five have non-zero apex curvature $\kappa$. The symbols are used for plotting in later figures.
	}
	\label{figshapes}
\end{figure}
Closed surfaces with sphere topology ($\chi=2$) are appealing due to their simplicity; but curved surfaces with a boundary ($\chi=1$) are arguably more commonplace, and pose qualitatively different problems. Such shapes necessitate a net charge $Q=6$, or (at least) six $q=+1$ disclinations. On nearly flat surfaces, these charges ideally occupy positions at the boundary of the structure where they do not contribute to mechanical energy, as the boundary shape can accommodate them without distortion (see Fig.~\ref{figshapes}a). For large enough curvature, it becomes favorable for at least one defect to migrate away from the boundary. Isolated instances of this transition have been analyzed for specific shapes such as spherical caps and paraboloids \cite{giomi2007crystalline,li2019,azadi2016neutral,azadi2014emergent}, including numerical approaches to this global optimization task in a class of problems considered $NP$ hard \cite{giomi2007crystalline}. 

It is desirable to formulate general intuitive criteria for this prototypical transition: what constitutes ``large enough curvature"? Anecdotally, it has been suggested that defects migrate to positions of large local $K_G$ \cite{vitelli2006crystallography}, but counterexamples to this rule are easily found. Recent literature instead proposes that a critical value of the integrated Gaussian curvature $\Omega=\int K_G \mathrm{d}A/(\pi/3)$ needs to be exceeded for defect migration into the bulk of the surface. However, significantly different critical values have been suggested for different experimental \cite{irvine2010pleats} and numerical systems \cite{giomi2007crystalline,azadi2016neutral,li2019}, casting doubt on the generality and accuracy of this criterion.

In the present work, we develop a new, physically motivated criterion that proves quantitatively accurate across a large class of shapes. We model defects as isolated singularities within a continuum linear elastic theory framework \cite{bowick2000interacting,nelson1995defects,bowick2001statistical}. Our focus is on a fixed surface geometry, excluding effects like buckling or faceting.

Prototypically, we explore the transition on arbitrary rotationally symmetric bounded surfaces from an energetically favorable defect-free state (i.e., all six disclinations are located at the boundary) to one with a single disclination at the center of the surface (Fig.~\ref{figshapes}a). While the presence of dislocations can strongly affect such transitions \cite{azadi2014emergent,azadi2016neutral}, for large defect core energies dislocation distributions are prohibitively expensive energetically, allowing us to focus on single-disclination energetics to study the onset of crystalline disorder \cite{nelson2002defects}.

We consider crystal lattices on smooth bounded surfaces in $\mathbb{R}^3$, and develop a formalism for elastic energy closely modeled on \cite{giomi2007crystalline} and \cite{bowick2009two}, of which we will only detail our variations and improvements (see the Supplemental Material \cite{suppl} for the complete derivations). The surfaces of revolution have the parametrization 
$Z=f(r)$. Generic surfaces have finite center curvature ($f''(0)\equiv\kappa\neq 0$), although we will also consider surfaces of higher-order flatness below. 
Surfaces have finite extent $0\leq r \leq r_{b}$; while we will consider shape families of both intrinsically finite extent (e.g. spheroids) and infinite extent (e.g. hyperboloids), we restrict ourselves to surfaces with unique $Z$ values, i.e., ``cap" shapes. The determinant $g$ of the metric tensor, Gaussian curvature $K_G$, and its integral $\Omega$ are then
\begin{subequations}
\begin{align}
\sqrt{g} = & r \sqrt{1+f'(r)^2},\quad K_G= \frac{f'(r)f''(r)}{r(1+f'(r)^2)^2}\label{rotsurface},\\
\Omega = & \,6\left(1-1/\sqrt{1+f'(r_b)^2}\right)\,.\label{omega}
\end{align}
\end{subequations}

\begin{figure}
	\centering
		\includegraphics[width=0.45\textwidth]{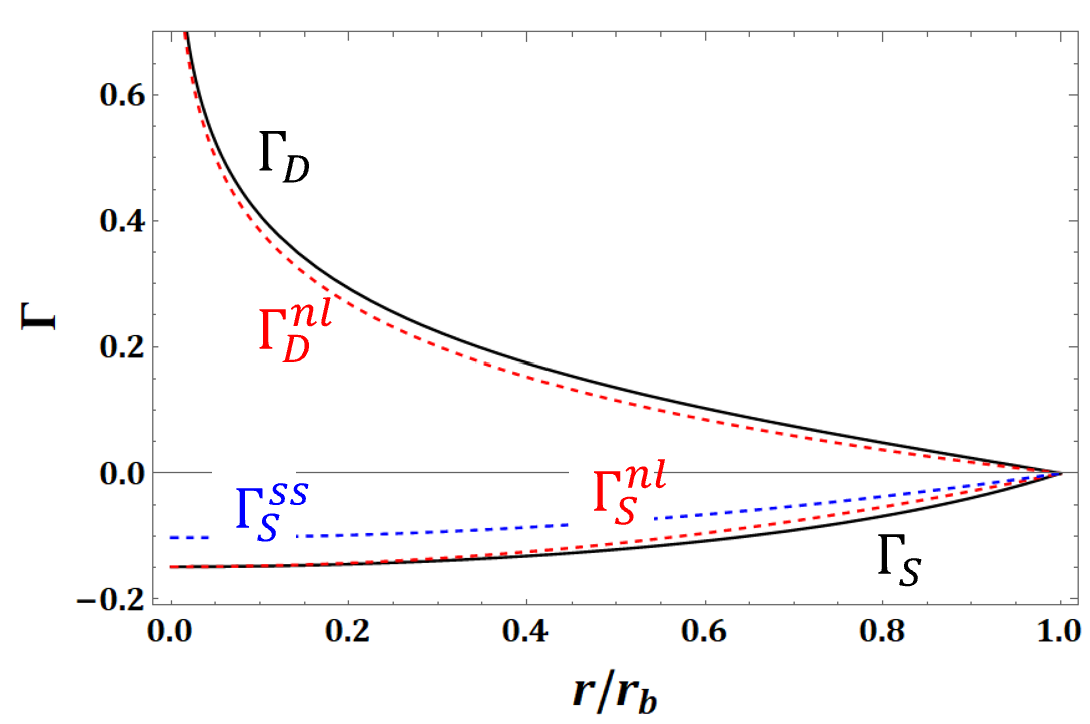}
	\caption{Comparison of $r$-dependent isotropic stress terms for the prototypical example of an oblate spheroid with $\kappa=0.8$. Black solid lines: exact expressions; blue dashed:  small-slope approximation; red dashed: non-local approximation}\label{gamma_ss_nl}
\end{figure}

With the free energy of the defect-free lattice and the defect core energy fixed, any minimization of energy is governed by $F_{el}$, the elastic energy associated with defect-surface interaction. Focusing on stress-free boundaries, the difference of $F_{el}$ for a configuration with only boundary defects (defect position $r_D=r_{b}$) and $F_{el}$ for one defect at the apex ($r_D=0$) is more explicitly written in terms of the traces of the stress tensor $\Gamma$,
\begin{align}
\Delta F_{el}(r_b)&\equiv F_{el}(0)-F_{el}(r_{b})\nonumber \\
    &=\pi \int\displaylimits_0^{r_{b}}  \left(\Gamma^2(r,0)-\Gamma^2(r,r_{b})\right) \sqrt{g}\,\mathrm{d} r,\label{Fel}
\end{align}
where $\Gamma(r,r_D)$ has four contributions 
\begin{align}
\Gamma(r,r_D) = -\Gamma_D(r,r_D) -\Gamma_S(r) + U_D(r_D) + U_K,\label{gamma2}
\end{align}
given by
\begin{subequations}
	\begin{align}
    \Gamma_D(r,0) =& - \frac{q}{6} \log \varrho(r),\qquad \Gamma_D(r,r_{b})=0,\label{gammad}\\
	\Gamma_S(r) =&\log \left(\varrho(r)\frac{r_{b}}{r}\right),\label{gammas}\\
	\varrho(r)=&\exp\left(- \int\displaylimits_r^{r_{b}}\frac{\sqrt{1+f'(r_1)^2}}{r_1} \mathrm{d}r_1\right)\, ,\label{rho}
	\end{align}\label{gamma}\noindent
\end{subequations}
representing exact functional forms, though the integral in \eqref{rho} may not have closed form. Here we have used integration by parts to simplify \eqref{gammas} further from \cite{giomi2007crystalline} (cf.\ \cite{suppl}). Energies have been made dimensionless by the Young's modulus for the planar crystal, and for the single-disclination case we will use $q=+1$ in the following. $\varrho(r)$ is the radius of conformal mapping of the surface onto the unit disk. \eqref{gammad} represents the contribution of the disclination while \eqref{gammas} captures the screening effect of Gaussian curvature. The generic shape of these functions is illustrated in Fig.~\ref{gamma_ss_nl}. 
Balancing $\Gamma_D$ and $\Gamma_S$ represents partial defect-charge compensation by local Gaussian curvature. The terms $U_D, U_K$ are determined by the boundary conditions and represent further energy compensation by integral terms over $\Gamma_D, \Gamma_S$. 
In our radially isotropic case with stress-free boundaries \cite{giomi2007crystalline}, $U_D(0)$ and $U_K$ are the straight averages
\begin{align}
    U_D(0)&=\frac{1}{A}\int \Gamma_D(r,0)\,\mathrm{d}A,\,\, U_K=\frac{1}{A}\int \Gamma_S(r)\, \mathrm{d}A,\label{Usymm}
\end{align}
with the surface area $A=2\pi\int \sqrt{g} \, \mathrm{d}r$.
We define the critical radius $r_{b}=r_c$ as the extent (or cap coverage) at which a center disclination becomes favorable, i.e., $\Delta F_{el}(r_c)=0$. Using \eqref{gamma2}, \eqref{gamma} and \eqref{Usymm} in \eqref{Fel} yields 
\begin{align}
    \Delta F_{el}(r_c)\equiv\int\displaylimits_0^{r_{c}} &\Gamma_D(r,0) (\Gamma_D(r,0) + 2\Gamma_S(r))\sqrt{g}\, \mathrm{d}r\nonumber\\&-U_D(0)(U_D(0)+2U_K)A=0, \label{delfel0}\noindent
\end{align}
as our rigorous criterion for transition of a disclination defect from a boundary position to the apex.

While (\ref{delfel0}) can be solved numerically, this is a rather opaque procedure and it is desirable to find an approximate criterion for the location of the transition that directly relates to surface shape. We compare two  approaches: (1) a local small slope approximation around $r=0$ used previously e.g. in \cite{azadi2016neutral}, and (2) a non-local approximation leading to a new, more accurate and widely applicable criterion for transition.  

\begin{figure*}
	\centering
\includegraphics[width=\textwidth]{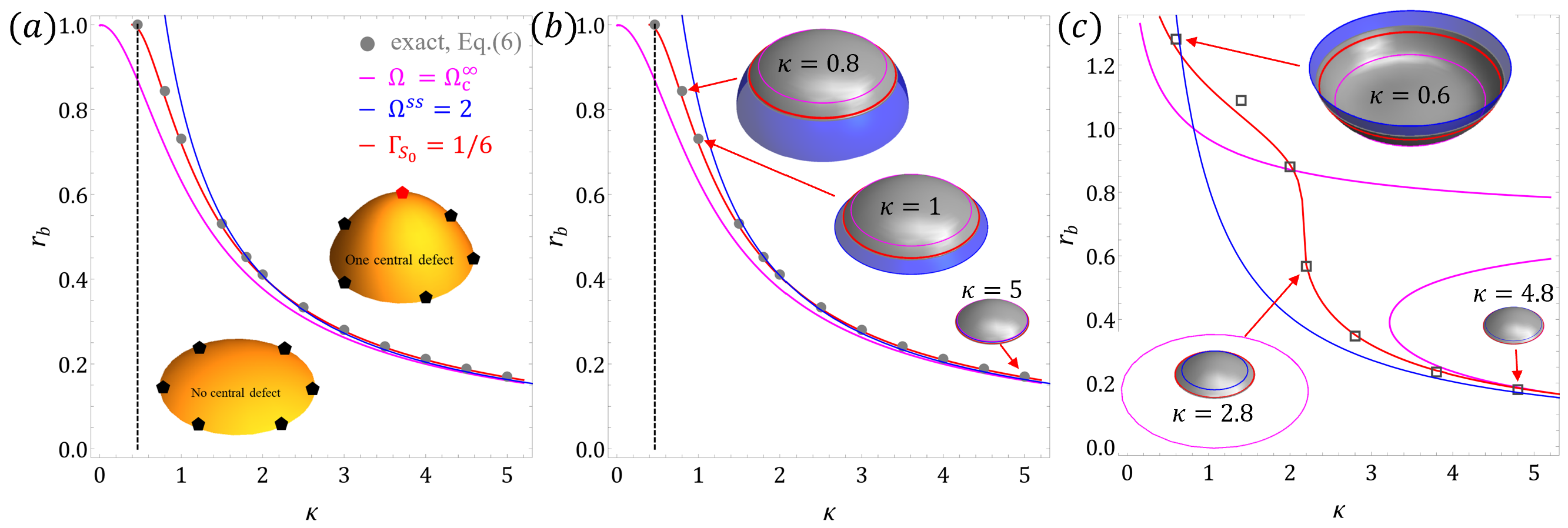}
	\caption{Cap extent at transition as a function of $\kappa$. Gray squares are numerically obtained critical points $r_c(\kappa)$ from \eqref{delfel0}; the solid blue lines result from the small-slope criterion \eqref{omsscrit}; the solid magenta lines are lines of constant $\Omega=\Omega_c^\infty$. The solid red lines are given by the criterion $\Gamma_{S_0}=1/6$ and are in excellent agreement with the rigorous transition points. (a) spheroids; the dashed vertical black line marks the minimal $\kappa_{c,min}$ for which transitions are present; 
	(b) visualization of spheroidal shapes at transition for various $\kappa$: oblate shapes bulge considerably while prolate shapes are very flat. Gray caps mark the exact extent, while the transition shapes determined from the approximate criteria are indicated in blue and magenta; red lines show the extent predicted from $\Gamma_{S_0}=1/6$;
	(c) transition lines for the $f(r)=\frac{\kappa}{3}(1-r^2+r^4)^{3/2}$ ``sombrero": this surface has a sign change in the boundary slope $s_b$ as $\kappa_c$ is increased along the transition boundary -- indicated by a sharp drop in $r_c$ around $\kappa_c\approx 2$. The $\Omega=const.$ criterion obtains more than one root for $\kappa_c \gtrsim 3$.
	}\label{spheroid_contour}
\end{figure*}

In the small-slope approach, the functions defined above are locally Taylor expanded around $r=0$ to leading order, requiring both $r\ll 1$ and $r_{b}\ll 1$, so that
\begin{subequations}
\begin{align}
    \Gamma_{S}^{ss}(r) &= \Gamma_S^{ss}(0)  \left(1-\frac{r^2}{r_{b}^2}\right),\label{ssapp}\\
    \Gamma_{D}^{ss}(r,0) &= -\frac{1}{6} \log \left(\frac{r}{r_{b}}\right) ,\quad \sqrt{g}^{ss} =r,\label{gammadss}
\end{align}
\end{subequations}
where $\Gamma_S^{ss}(0)=-\frac{1}{4} \kappa^2 r_{b}^2$ is the small-slope expansion of the full $\Gamma_S(0)$. 
Inserting into \eqref{Usymm} and  \eqref{delfel0} gives a straightforward polynomial equation in $\kappa$ and $r_{c}$, whose solution yields a criterion for critical extent as \cite{azadi2016neutral}
\begin{align}
    r_{c}=\frac{\sqrt{2/3}}{\kappa}\quad \text{or} \quad  \Omega^{ss} = 3\kappa^2r_{c}^2 =2 \,.
    \label{omsscrit}
\end{align}
This formalism thus suggests that the transitional surface shape is indeed given by a universal (leading-order) value of integrated Gaussian curvature. However, different studies have determined values of this quantity that differ by more than a factor of two \cite{irvine2010pleats,azadi2016neutral,giomi2007crystalline} depending on the surface considered.

An inherent flaw of the small-slope approximation is apparent when comparing the function $\Gamma_S^{ss}$ to its exact version: not even the value at the expansion point, $\Gamma_S(0)$, is accurately reproduced (see Fig.~\ref{gamma_ss_nl}, which shows a representative case). We remedy this problem by requiring this  matching to hold, replacing the local quantity $\Gamma_S^{ss}(0)$ by the exact value, so that (\ref{ssapp}) is modified to
\begin{align}
    \Gamma_{S}^{nl}(r) &= \Gamma_S(0) \left(1-\frac{r^2}{r_{b}^2}\right),\label{asyleading0}
\end{align}
with the rigorous expression
\begin{align}
   \Gamma_S(0)\equiv& \int\displaylimits_0^{r_{b}} K_G(r) \log\varrho(r) \sqrt{g}\,\mathrm{d}r\label{gams0}
\end{align}
i.e., the trace of the full background stress tensor at the apex, a non-local quantity which represents an integrated Gaussian curvature {\em weighted} by the $\log \varrho$ singularity characteristic of the Green's function of the problem \cite{suppl}, and thus of the local stress due to the defect $\Gamma_D$.
For now, we leave the other approximations unchanged, $\Gamma_{D}^{nl}=\Gamma_{D}^{ss}$ and $\sqrt{g}^{nl}=\sqrt{g}^{ss}$.

When using \eqref{gammadss}, \eqref{asyleading0} and \eqref{Usymm} in the exact \eqref{delfel0}, the integration from $r=0$ to $r=r_{c}$ is again straightforward and results in a simple prediction for $\Gamma_{S}(0)$:
\begin{align}
    \Gamma_{S_0}\equiv -\Gamma_{S}(0)= \frac{1}{6}\label{gams0crit}
\end{align}
The LHS from (\ref{gams0}) is an implicit equation in the cap extent $r_c$ and the shape parameters (e.g.\ $\kappa$). The new criterion (\ref{gams0crit}) is as conceptually simple as (\ref{omsscrit}), but recognizes that the Gaussian curvature in each point contributes to stress relief of a central disclination with different weight. 

We now examine how this simple condition on $\Gamma_{S_0}$ performs. Going beyond evaluations for individual surface shapes (cf.\ \cite{azadi2016neutral,li2019}, \cite{giomi2007crystalline}),
we investigate the location of the prototypical disclination transition in entire families of  surfaces of revolution (illustrated in Fig~\ref{figshapes}b) with various distributions of curvature.

In the family of spheroids, $f(r) = \kappa  \sqrt{1-r^2}$, prolate ($\kappa>1$) shapes have a Gaussian curvature maximum at the apex, while oblate ($\kappa<1$) spheroids have maximum $K_G$ at the boundary. Figure~\ref{spheroid_contour}(a) plots the critical values $r_b=r_c$ as a function of $\kappa$. The symbols result from numerical evaluation of the exact covariant criterion (\ref{delfel0}) and show that both prolate and oblate shapes display disclination migration as long as $\kappa>\kappa_{c,min}\approx 0.465$. Thus, maximum local curvature at the (apex) disclination position is not sufficient to determine that position. For the particular case of a spherical cap ($\kappa=1$), we note that our results agree with the findings of \cite{li2019}, and the exact covariant criterion \eqref{delfel0} results in $r_c\approx 0.73$.

Plotting the small-slope formula (\ref{omsscrit}) as an $r_c(\kappa)$ contour in Fig.~\ref{spheroid_contour}a shows that it is relatively accurate at large $\kappa$, but exhibits strong deviations for $\kappa\lesssim 1$, failing to predict even the existence of a transition for the most oblate shapes. Testing the hypothesis of constant integrated  Gaussian curvature, we also plot a contour of constant $\Omega=\Omega_c^\infty\approx 1.35$ evaluated from (\ref{omega}). This value is obtained by matching the threshold for this exact $\Omega$ criterion to the small-slope value as $\kappa\to\infty$. 
Significant errors are again apparent for smaller $\kappa \lesssim 2$, this time underestimating $r_c$ so that the true transition shape requires ``overcharging" of the surface as observed in previous studies \cite{irvine2010pleats,li2019}. Clearly, $\Omega$ is not a constant at the transition of disclination migration (at $\kappa_{c,min}$, $\Omega_c^\infty$ is about a factor of four smaller than $\Omega$ at the true transition point). 

By contrast, the new indicator for critical shapes $\Gamma_{s_0}=1/6$ derived above yields a transition curve in excellent agreement with the rigorous transition threshold for all spheroidal caps, cf.\ Fig.~\ref{spheroid_contour}(a,b). The $\Omega$ and $\Omega^{ss}$ criteria can fail much more severely for surfaces with strong variation of $K_G$, such as the family of ``sombrero" shapes depicted on the lower left of Fig.~\ref{figshapes}. Nevertheless, the rigorous transition is very well captured by the $\Gamma_0$ criterion, see
Fig.~\ref{spheroid_contour}(c).

In all cases, the criteria based on integrated Gaussian curvature perform poorly for $\kappa\sim 1$.
We note that in applications in nature and technology, these cases are the most practically relevant: Fig.~\ref{spheroid_contour}(b) shows that transition shapes for large $\kappa$ are so small that they are nearly flat caps, while $\kappa\sim 1$ shapes are ``bulgy" (have an aspect ratio of height to diameter near one). These are the shapes that non-trivially reconcile significant curvature and crystalline order, such as in an insect eye that needs to bulge (for field of view) while having ordered facets (for accurate processing of the visual information). 

\begin{figure}[t]
	\centering
\includegraphics[width=0.5\textwidth]{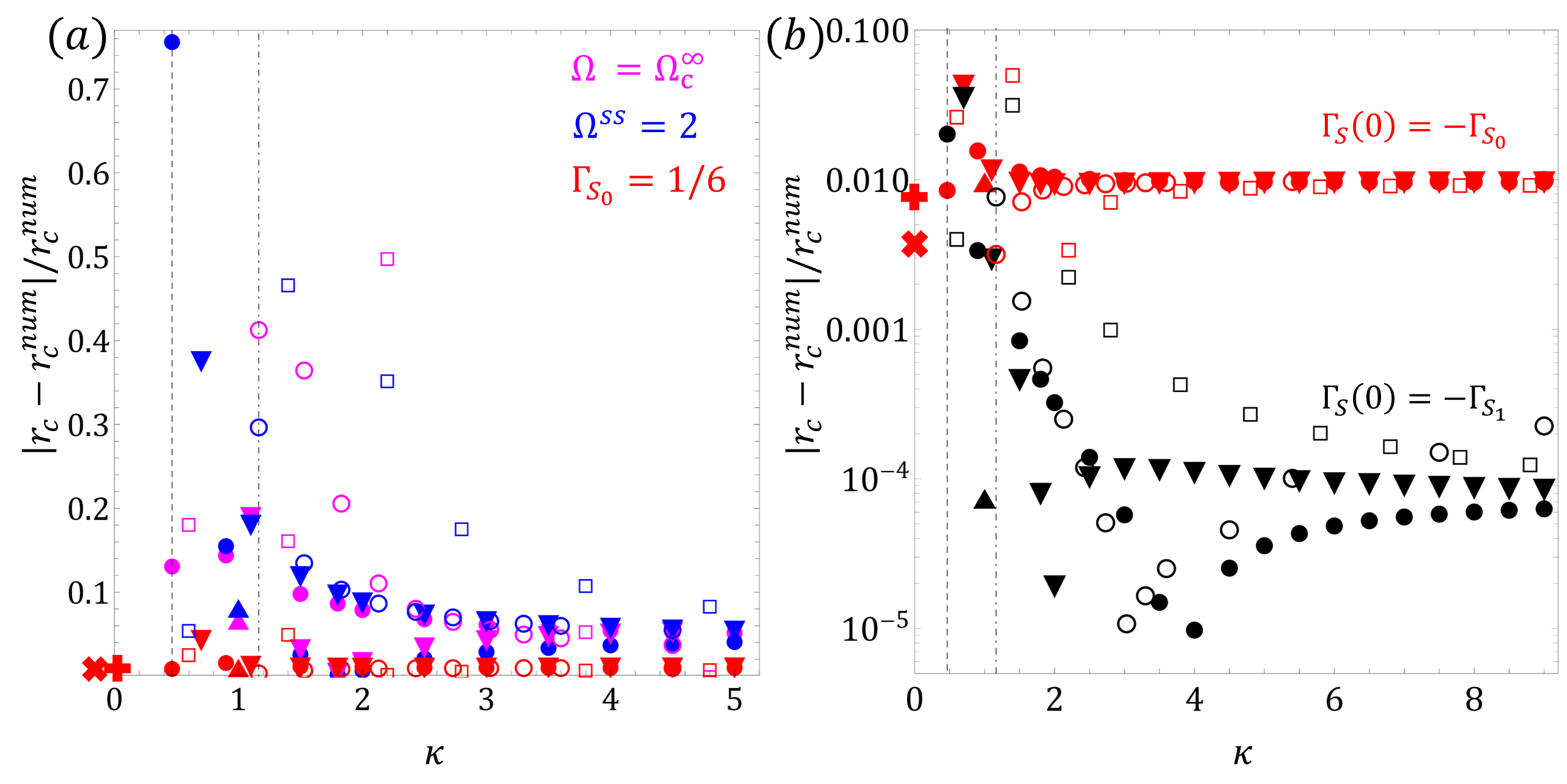}\label{omctrans}
\caption{Relative errors in $r_c(\kappa)$ using different transition predictors (refer to Fig.~\ref{figshapes}(b) for a key to symbols). 
Dashed and dot-dashed vertical lines indicate limiting $\kappa_{c,min}$ for spheroids (filled circles) and ``bell-shaped" surfaces (open circles), respectively. The two shapes of zero apex curvature from  Fig.~\ref{figshapes}(b) are represented by symbols at $\kappa=0$. (a) Blue symbols use \eqref{omsscrit}, magenta use constant $\Omega$, red symbols use $\Gamma_{S_0}= 1/6$. (b) Red symbols: same as (a) on a log scale; black symbols: improved transition criterion using $\Gamma_{S_1}$.  }\label{transitiongoodness}
\end{figure}

These findings hold true for all the shape families in Fig.~\ref{figshapes}(b). Figure~\ref{transitiongoodness} quantifies the relative errors of $r_c(\kappa)$ for the three different criteria. Regardless of whether the parametrization allows for infinite caps (e.g.\ hyperboloids) and of whether the Gaussian curvature changes sign or not, the criterion of $\Gamma_{S_0}=1/6$ uniformly outperforms those based on integrated $K_G$, with most errors below $1\%$. We note that for a paraboloid (upward triangles at $\kappa=1$ in Fig.~\ref{transitiongoodness}), we find $r_c\approx 0.856$, while \cite{giomi2007crystalline} obtained $r_c\approx 1.5$ as a consequence of neglecting $U_D(0)$ in the energy evaluation.
It is apparent from Figs.~\ref{spheroid_contour} and \ref{transitiongoodness} that the transition positions (and errors) have a common asymptote for $\kappa\to\infty$. By Taylor-expanding the implicit equation (\ref{gams0crit}) for small $r_{c}$ and large $\kappa$, we obtain the asymptotic form  of the transition line for any surface of revolution with large apex curvature:
\begin{align}
    r_c(\kappa) \approx \frac{0.848}{\kappa} - \frac{0.146}{\kappa^3} +\frac{0.02}{\kappa^5} + \mathcal{O}\left(\frac{1}{\kappa^7}\right).
    \label{largekappa}
\end{align}
Note that the leading order of the above equation is not equivalent to the condition $r_c = \sqrt{2/3}/\kappa_c\approx 0.816/\kappa_c$  from the small-slope approximation (\ref{omsscrit}). The new criterion improves even this leading order, and furthermore provides higher-order corrections. 

Perhaps surprisingly, the $\Gamma_{S_0}$ criterion remains as accurate for surfaces with $\kappa=0$ that do not share the asymptote (\ref{largekappa}). Two examples (the last two shapes in Fig.~\ref{figshapes}) are indicated in Fig.~\ref{transitiongoodness}. This demonstrates the non-perturbative character of (\ref{gams0crit}).

The criterion based on weighted integrated Gaussian curvature is not only more accurate, but more systematic: In Fig.~\ref{transitiongoodness}(b), we show that the next-order expansion $\Gamma_{S_1}$ of $\Gamma_S(0)$  provides a further dramatic improvement of relative errors across all shapes, particularly for larger $\kappa$.
For these results, the non-local computation of $\Gamma_S(0)$ was maintained, while further terms of relative order $(r/r_{b})^2$ were included in (\ref{asyleading0}) and determined by enforcing a match to derivatives at $r=r_{b}$. Explicitly, this leads to 
\begin{subequations}
\begin{align}
    \Gamma_{S}^{nl}(r) =& \Gamma_S(0) \left(1-\frac{r^2}{r_{b}^2}\right),\\
    \Gamma_{D}^{nl}(r,0)=& -\frac{1}{6}\left[ \log \left(\frac{r}{r_{b}}\right)+\frac{\left(1-\sqrt{1+s_b^2} \right)}{2}\left(1-\frac{r^2}{r_{b}^2}\right)\right], \\
    \sqrt{g}^{nl}(r) =& r + \left( r_{b}\sqrt{1+s_b^2} - r_{b}\right) \frac{r^3}{r_{b}^3}\,,
\end{align}
\label{asyleading1}\noindent
\end{subequations}
defining the slope at the boundary as $s_b\equiv f'(r_{b})$. 
Inserting into (\ref{delfel0}) and integrating results in 
\begin{align}
    &\Gamma_{S_1} = -\Gamma_{S}(0)= \\
    &\frac{70+2s_b^4+74\sqrt{1+s_b^2}+s_b^2\left(41+12\sqrt{1+s_b^2}\right)}{48\left(8+10\sqrt{1+s_b^2}+s_b^2\left(5+\sqrt{1+s_b^2}\right)\right)}\nonumber\\
    &\approx \frac{1}{6}+\frac{1}{432}s_b^2+ \dots \,.\label{gams1crit}
\end{align}
We compare the $\Gamma_{S_0}$ and $\Gamma_{S_1}$ criteria in Fig~\ref{transitiongoodness}(b). The ${\cal O}(s_b^2)$ term in (\ref{gams1crit}) has a very small prefactor, making the difference between $\Gamma_{S_0}$ and $\Gamma_{S_1}$ slight even for moderate $s_b$. The transition shape of surfaces with large $\kappa$ is such that $s_b^2\approx \kappa^2 r_{b}^2 \to 0.848$, see (\ref{largekappa}), which allows quantification of the second term of (\ref{gams1crit}) in that limit. We obtain $s_b^2/432\to 0.0099\dots$, explaining the magnitude of the universal error asymptote of $\Gamma_{S_0}$ as $\kappa\to\infty$. Thus, the error of the non-local criterion for transition is controlled and predictable.

The transition criterion introduced here argues that the primary determinant of the onset of disclination disorder on an open crystalline surface is neither just the local Gaussian curvature nor its straight integral $\Omega$, but that the values of $K_G$ everywhere contribute proportional to the local isotropic stress due to the disclination (cf.\ the functional form of $\Gamma_D$ in (\ref{gammad})). This quantity $\Gamma_{S_0}$ is as easy to calculate as $\Omega$ and has a universal value at transition for all shape families investigated here. It is an accurate predictor particularly for surfaces of moderate central curvature that represent practically relevant, significantly ``bulging" shapes that still maintain crystalline order. Further work will be devoted to generalizing this criterion to anisotropic surfaces and changes in boundary conditions, as well as to the prediction of the first- or second-order character of the transitions.  

We are grateful to Mark Bowick, Paul Chaikin, and Greg Grason for insightful discussions. SA acknowledges support by the NSF under grant $\#$ 1504301.

\bibliographystyle{apsrev4-2}
\bibliography{ms}

\end{document}


\sloppy 

\title{Supplemental Material: A simple, general criterion for onset of disclination disorder on curved surfaces}
\author{Siddhansh Agarwal}
\author{Sascha Hilgenfeldt}%
\affiliation{Mechanical Science and Engineering, University of Illinois, Urbana-Champaign, Illinois 61801, USA}%
\maketitle
\section{Exact covariant formalism}
One way to compute stress on curved elastic surfaces is to use the
Airy stress function $\chi$. It solves the following inhomogeneous biharmonic equation \cite{giomi2007crystalline,bowick2009two}:
\begin{align}
\Delta^2 \chi(\mathbf{x},\mathbf{x}_D) =Y_0 q_T(\mathbf{x},\mathbf{x}_D) \label{biharm},
\end{align}
where $q_T(\mathbf{x},\mathbf{x}_D) = \frac{\pi}{3} \delta(\mathbf{x},\mathbf{x}_D) -K_G(\mathbf{x})$ represents both the singular contributions of disclinations and the continuous contribution of surface shape (Gaussian curvature); $Y_0$ is the 2D Young's modulus of the surface. We impose no-stress boundary conditions on the domain $\mathbb{P}$:
\begin{subequations}
	\begin{align}
	\chi(\mathbf{x},\mathbf{x}_D) = 0, \quad \mathbf{x}\in \partial \mathbb{P},
	\end{align}
	\begin{align}
	\nu_i \nabla^i \chi(\mathbf{x},\mathbf{x}_D) = 0,\quad \mathbf{x}\in \partial \mathbb{P} \label{neumann}
	\end{align}
\end{subequations}
with normal $\nu_i$. The solution of \eqref{biharm} will then be
\begin{align}
\chi(\mathbf{x},\mathbf{x}_D) = \int \mathrm{d} \mathbf{y} \,G_L (\mathbf{x},\mathbf{y}) \Gamma(\mathbf{y},\mathbf{x}_D),
\end{align}
where $G_L (\mathbf{x},\mathbf{y})$ is the Green's function of the covariant Laplace operator on $\mathbb{P}$ with Dirichlet boundary conditions
\begin{subequations}
	\begin{align}
	\Delta G_L(\mathbf{x},\mathbf{.}) = \delta(\mathbf{x},\mathbf{.}), \quad \mathbf{x}\in \mathbb{P},
	\end{align}
	\begin{align}
	G_L(\mathbf{x},\mathbf{.}) = 0,\quad \mathbf{x}\in \partial \mathbb{P},
	\end{align}
\end{subequations}\label{Gl}
and $\Gamma(\mathbf{x},\mathbf{x}_D)=\Delta\chi(\mathbf{x},\mathbf{x}_D)$ is the solution of the Poisson problem
\begin{align}
\Delta\Gamma(\mathbf{x},\mathbf{x}_D) = Y_0 q_T(\mathbf{x},\mathbf{x}_D)\,.
\end{align}
This solution can be expressed formally as
\begin{align}
\Gamma(\mathbf{x},\mathbf{x}_D) &=Y_0 \int q_T(\mathbf{y},\mathbf{y}_D)G_L(\mathbf{x},\mathbf{y}) \mathrm{d} \mathbf{y} \nonumber\\
&= -\Gamma_D(\mathbf{x},\mathbf{x}_D) -\Gamma_S(\mathbf{x}) + U(\mathbf{x},\mathbf{x}_D),\label{gammaxapp}
\end{align}
where
\begin{subequations}
	\begin{alignat}{4}
    \Gamma_D(\mathbf{x},\mathbf{x}_D) &= - \frac{\pi}{3}Y_0 G_L(\mathbf{x},\mathbf{x}_D),\label{gammadapp}\\
	\Gamma_S(\mathbf{x}) &=Y_0 \int K_G(\mathbf{y}) G_L(\mathbf{x},\mathbf{y}) \mathrm{d} \mathbf{y}, 
	\end{alignat}
\end{subequations}
and $U(\mathbf{x},\mathbf{x}_D)$ is a harmonic function on $\mathbb{P}$ that enforces the Neumann boundary conditions. The first term of \eqref{gammaxapp} represents the bare contribution of disclinations while the second term captures the screening effect of Gaussian curvature. In this paper we restrict ourselves to allowing only one disclination to migrate from the boundary to the apex of the manifold. For symmetric $\Gamma_S$ and $\Gamma_D$, i.e. isotropic surface and defects decorated at centre or boundary, the harmonic function can be computed by directly applying the boundary condition \eqref{neumann} in \eqref{gammaxapp} and this results in 
\begin{align}
    U=\frac{1}{A}\int \left(\Gamma_D(\mathbf{x},\mathbf{x}_D)+\Gamma_S(\mathbf{x})\right)\,\mathrm{d}\mathbf{x},\label{Usymm}
\end{align}
where the surface area $A=\int \mathrm{d}\mathbf{x}$. The elastic energy for a stress-free boundary can be expressed only in terms of the isotropic stress tensor $\Gamma$ as
\begin{align}
     F_{el}(\mathbf{x}_D)=\frac{1}{2Y_0} \int  \Gamma^2(\mathbf{x},\mathbf{x}_D)\,\mathrm{d} \mathbf{x},\label{Fel}
\end{align}
The Green's function satisfying \eqref{Gl} is computed explicitly by conformally mapping the surface $\mathbb{P}$ onto the unit disk of the complex plane where the Green's function is known:
\begin{align}
G_L (\mathbf{x},\mathbf{y}) = \frac{1}{2\pi} \log \left|\frac{z(\mathbf{x})-z(\mathbf{y})}{1-z(\mathbf{x}) \overline{z(\mathbf{y})}}\right|,\label{Glapp}
\end{align}
where $z(\mathbf{x}) = \varrho e^{i\phi}$, a point in the unit disk, is the image of a point on the surface $\mathbb{P}$ under the conformal mapping. The Green's function vanishes when the disclination is located at the boundary. For a surface $X(r,\phi)$ with first fundamental form $E=\partial X/\partial r \cdot \partial X/\partial r$, $F=\partial X/\partial r \cdot \partial X/\partial \phi$ and $G=\partial X/\partial \phi \cdot \partial X/\partial \phi$, the metric of the surface is
\begin{align}
\mathrm{d}s^2 = E \mathrm{d}r^2 + 2F \mathrm{d}r\mathrm{d}\phi +G \mathrm{d}\phi^2 \label{metricX}\,,
\end{align}
whereas the unit disk has the metric
\begin{align}
\mathrm{d}s^2 = w(z)\left(\mathrm{d}\varrho^2 + \varrho^2 \mathrm{d}\phi^2\right) \label{metricdisk}\,.
\end{align}
where $w(z)$ is a positive conformal weight. The remaining task is now to find the conformal factor $w(z)$ and the conformal radius $\varrho(r)$ by equating the two metrics; these can be explicitly obtained for many rotationally symmetric surfaces but in general, may not be analytically computable.

Taking the two image points on the unit disk as $z(r,\phi)=\varrho_x(r)e^{i\phi}$ and $\zeta(r',\phi')=\varrho_y(r')e^{i\phi'}$, the contribution due to the background Gaussian curvature is split into two parts $\Gamma_S(\mathbf{x}) = \Gamma_{S,1}(\mathbf{x}) - \Gamma_{S,2}(\mathbf{x})$, where
\begin{subequations}
	\begin{align}
	\Gamma_{S,1}(\mathbf{x}) &= \frac{Y_0}{2\pi} \int \mathrm{d} \phi' \, \mathrm{d}r' \, \sqrt{g} K(r')\log|z-\zeta|,\\
	\Gamma_{S,2}(\mathbf{x}) &= \frac{Y_0}{2\pi} \int \mathrm{d} \phi' \, \mathrm{d}r' \, \sqrt{g} K(r')\log|1-z\overline{\zeta}| ,
	\end{align}\label{gammas12}
\end{subequations}
are evaluated analytically for the specific surfaces considered in this paper. 

For rotationally symmetric surfaces parametrized by $X=r \cos\phi,\, Y=r \sin\phi,\, Z=f(r)$ with the first fundamental form $E=1+f'(r)^{2}$, $F=0$, and $G=r^2$, the conformal distance $\varrho$ (on the unit disk) can be obtained by equating \eqref{metricX} and \eqref{metricdisk}, so that one obtains
\begin{subequations}
    \begin{align}
        w(\varrho)&=\frac{r^2}{\varrho^2},\\
        \frac{d\varrho}{dr} &=\mp \frac{\sqrt{1+f'(r)^2}}{r} \varrho. \label{varrhoode}
    \end{align}
\end{subequations}
This last ODE may be solved analytically with boundary conditions $\varrho(0)=0$ and $\varrho (r_{b})=1$, which yields
\begin{align}
\varrho (r) &= \exp \left(-\int_r^{r_{b}}\sqrt{E/G}\,\mathrm{d}r_1\right)=\exp\left(-\int_r^{r_{b}} \frac{\sqrt{1+f'(r_1)^2}}{r_1}\,dr_1\right),
\end{align}
Note that in the small-slope limit, i.e. $f'(r_{b})\ll 1$, the conformal distance $\varrho(r)=r/r_{b}$ which physically means that the manifold is a flat disk in this limit. $\Gamma_S$ can be further simplified using the expansions 
\begin{subequations}
\begin{align}
    \log |z-\zeta|&=\log \varrho_> - \sum_{n=1}^\infty\frac{1}{n}\left(\frac{\varrho_<}{\varrho_>}\right)^n \cos n(\phi-\phi')\\
    \log |1-z\overline{\zeta}|&= - \sum_{n=1}^\infty\frac{1}{n}\left(\varrho \varrho'\right)^n \cos n(\phi-\phi')
\end{align}
\end{subequations}
where $\varrho_>(\varrho_<)$ represents the largest (smallest) modulus between $z$ and $\zeta$. If $K_G$ and $\sqrt{g}$ are azimuthally symmetric then all the angular dependences in \eqref{gammas12} vanish so that we have
\begin{align}
    \Gamma_{S,1}(r) =& Y_0\log \varrho(r) \int_0^r  \frac{f'(r_1)f''(r_1)}{\left(1+f'(r_1)^2\right)^{3/2}}\mathrm{d}r_1 \nonumber\\
    &+Y_0\int_r^{r_{b}}  \log \varrho(r_1) \frac{f'(r_1)f''(r_1)}{\left(1+f'(r_1)^2\right)^{3/2}}\mathrm{d}r_1,\label{gammas1app}\\
    \Gamma_{S,2}(r) &= 0.
\end{align}
The first integral in \eqref{gammas1app} is the integrated Gaussian curvature and can be analytically executed using the fact that $f'(0)=0$ (since by symmetry the slope at the apex is zero), while the second one may be integrated by parts to obtain
\begin{align}
    \Gamma_{S,1}(r) =& Y_0\log \varrho(r) \left(1 - \frac{1}{\sqrt{1+f'(r)^2}}\right) + Y_0\left[\log \varrho(r_1) \frac{-1}{\sqrt{1+f'(r_1)^2}}\right]_r^{r_{b}} -Y_0\int_r^{r_{b}}  \frac{1}{\varrho(r_1)}\frac{d\varrho(r_1)}{dr_1} \frac{-1}{\sqrt{1+f'(r_1)^2}}\mathrm{d}r_1\nonumber\\
    =& Y_0\log \varrho(r) + Y_0\int_r^{r_{b}} \frac{1}{r_1} \mathrm{d}r_1\nonumber\\
    =& Y_0\log \varrho(r) + Y_0\log \left(\frac{r_{b}}{r}\right) = Y_0\log\left(\varrho(r) \frac{r_{b}}{r}\right),
\end{align}
where we have used \eqref{varrhoode} and the fact that $\log \varrho(r_{b})=\log 1=0$. It is also evident that $\Gamma_S(0)$ is just the second integral in \eqref{gammas1app} so that 
\begin{align}
    \Gamma_{S}(0)&=Y_0\int_0^{r_{b}}  \log \varrho(r_1) \frac{f'(r_1)f''(r_1)}{\left(1+f'(r_1)^2\right)^{3/2}}\mathrm{d}r_1,\nonumber\\
   &=Y_0 \left[\int_0^{r_b} \log r \frac{f'(r)f''(r)}{\sqrt{1+f'(r)^2}} \mathrm{d} r +\log r_{b}\left(1-\sqrt{1+f'(r_{b})^2}\right)\right]
\end{align}
is a weighted integrated Gaussian curvature, expressed in terms of the parametrization $f(r)$. Additionally, for the symmetric case of a disclination positioned at the center, we have
\begin{align}
    \Gamma_D(r) = -\frac{Y_0}{6}\log \varrho(r)
\end{align}
which can be readily seen by setting $\mathbf{x}_D=0$ in \eqref{gammadapp} and \eqref{Glapp}. This demonstrates, for the normalization $Y_0=1$, the expressions for $\Gamma_S$ and $\Gamma_D$ in the main text. 

\section{Additional Transition shapes}

\begin{figure}[H]
	\centering
		\includegraphics[width=0.7\textwidth]{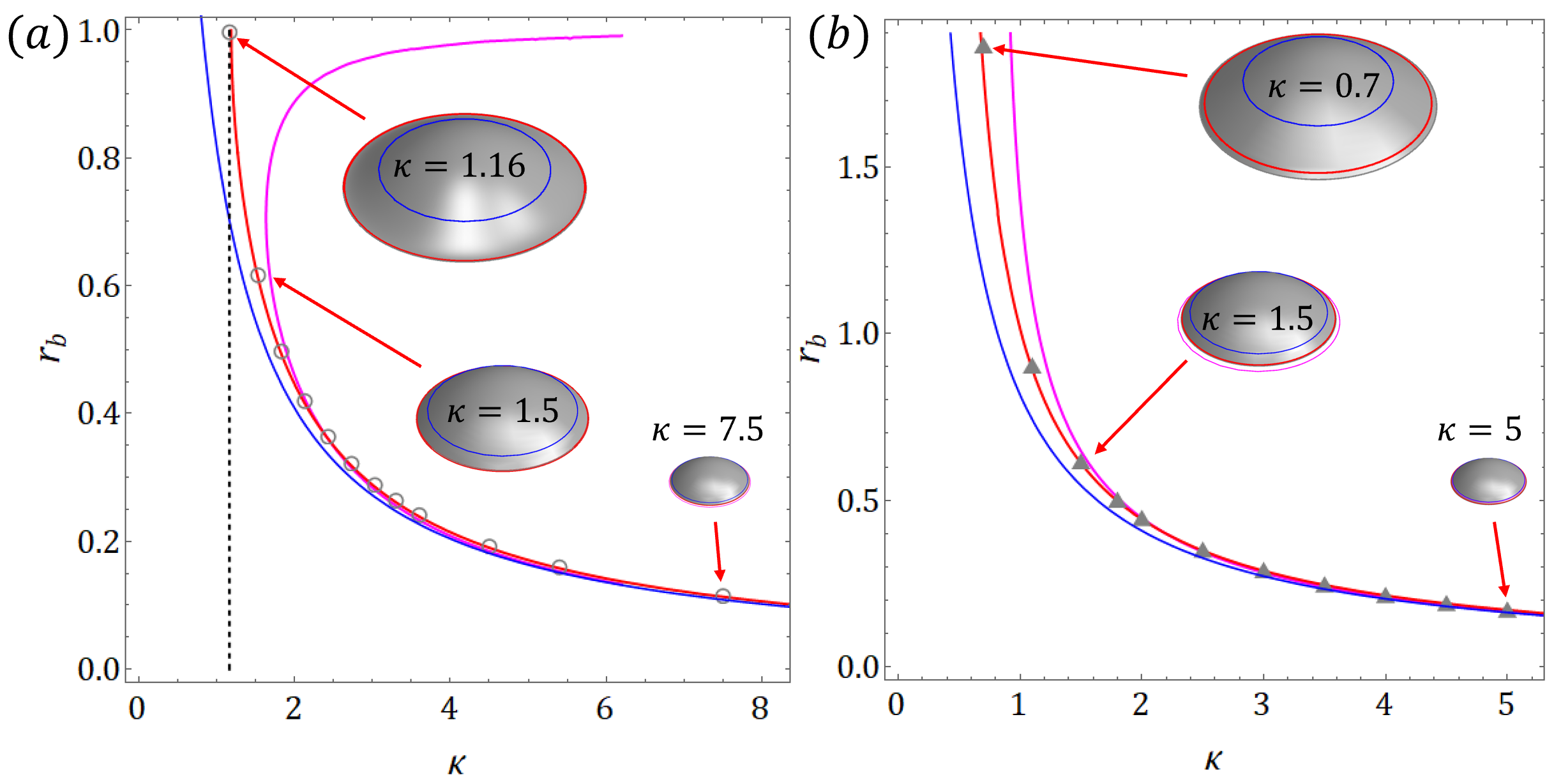}
	\caption{Visualization of shapes at transition for: (a) $f(r)=\frac{\kappa}{3}(1-r^2)^{3/2}$ and (b) $f(r)=\kappa\sqrt{(1+r^2)}$. Gray symbols are numerically obtained exact critical points; the solid blue line is the small-slope criterion $\Omega^{SS}=2$; the solid magenta line is a line of constant $\Omega=\Omega_{c,\infty}$ and the solid red line is the non-local criterion $\Gamma_{S_0}=1/6$. The dashed vertical black line indicates the minimal $\kappa_{c,min}$ for which transitions are observed. Gray caps illustrate the exact extent of caps for the $\kappa$ values indicated, while the transition shapes determined from the approximate criteria are shown in blue and magenta; red lines indicate the extent predicted from $\Gamma_{S_0}=1/6$}\label{figshapessuppl}
\end{figure}

We present additional evidence on the general applicability of the transition criterion $\Gamma_{S_0}=1/6$ by analyzing other shape families.
In Figure \ref{figshapessuppl}, we show how the different transition criteria perform for (a) ``bell"-shapes $f(r)=\frac{\kappa}{3}(1-r^2)^{3/2}$ and (b) hyperboloids $f(r)=\kappa\sqrt{(1+r^2)}$. A constant $\Omega$ criterion unphysically predicts no roots or double roots for some values of $\kappa_c$ while a small-slope constant $\Omega^{SS}$ criterion fails to accurately account for surface details particularly when $\kappa_c\sim 1$. Our $\Gamma_{S_0}=1/6$ criterion, on the other hand, performs exceedingly well for these shape families, as well as for all others that were tested.

\bibliography{ms}